\documentclass[twocolumn,psfig,epsfig,aps,prl,showpacs]{revtex4}
\usepackage{graphics}
\def\beq{\begin{equation}}
\def\eeq{\end{equation}}
\def\bea{\begin{eqnarray}}
\def\eea{\end{eqnarray}}
\def\lf{\left}
\def\rt{\right}

\def\tst{\textstyle}

\def\fno#1{Fig.~\ref{#1}}

\def\eno#1{Eq.~(\ref{#1})}

\def\etwo#1#2{Eqs.~(\ref{#1}) and (\ref{#2})}

\def\al{\alpha}

\def\gam{\gamma}
\def\dta{\delta}
\def\eps{\epsilon}
\def\tta{\theta}

\def\lam{\lambda}

\def\Dta{\Delta}

\def\ptl{\partial}

\def\hf{{1\over2}}
\def\tshf{\tst\hf}

\def\ham{{\cal H}}
\def\ket#1{|#1\rangle}

\def\tran#1#2{\langle#1|#2\rangle}

\def\mel#1#2#3{\langle#1|#2|#3\rangle}

\def\bJ{{\bf J}}

\def\xhat{{\bf{\hat x}}}

\def\zhat{{\bf{\hat z}}}
\def\nhat{{\bf{\hat n}}}

\def\Fe8{Fe$_8$}

\def\rtl{\sqrt{\lam}}

\def\baz{{\bar z}}

\def\bzi{{\bar z}_i}
\def\bzf{{\bar z}_f}
\def\Scl{S^{{\rm cl}}}
\def\emin{E_{{\rm min}}}

\begin{document}


\title{SU(2) instantons with boundary jumps and spin
tunneling in magnetic molecules}

\author{Ersin Ke\c{c}ecio\u{g}lu}
\author{Anupam Garg}
\email[e-mail address: ]{agarg@northwestern.edu}
\affiliation{Department of Physics and Astronomy, Northwestern University,
Evanston, Illinois 60208}

\date{\today}

\begin{abstract}
Coherent state path integrals are shown in general to contain instantons
with jumps at the boundaries, i.e., with boundary points lying outside
classical parameter or phase space. As an example, the magnetic molecule
\Fe8 is studied using a realistic Hamiltonian, and instanons with jumps are
shown to dominate beyond a certain external magnetic field. An approximate
formula is found for the fields where ground state tunneling is quenched
in this molecule.
\end{abstract}


\maketitle

The purpose of this paper is to discuss two problems, a specific
one and a general one. The specific problem concerns the
tunneling between ground Zeeman levels of certain magnetic
molecules \cite{werns}. The general problem concerns the nature
of tunneling paths, or instantons, in coherent state or phase space path
integrals \cite{lsaa}. As a rule, such paths can be found only if one
complexifies phase space, i.e., allows either the
momenta, or the coordinates, or both, to become complex.
However, in all previous studies of which we are aware,
the instantons always start and end in real phase
space, at the points corresponding to the classical energy minima.
The new instantons reported here, by contrast, do not even have
end points in real phase space. We refer to these as boundary jump
instantons. Such paths are a basic part of the formal structure of
coherent state and phase space path integrals \cite{Fadd,Klau}, but
there has never been a need to include them in tunneling problems,
as paths without jumps have always been available \cite{mis1}. This,
as we shall show, is not an accident. Boundary jump instantons are
analogous to extra saddle points in the method of steepest descents for
one-dimensional integrals, and like them, may or may not be relevant
in any given situation. But, one can not ignore them a priori. In this
paper we give the rules for finding these extra paths and their
contribution to tunneling, and apply them to our illustrative example.

Our work is part of the broader program of studying the semiclassical
limit of spin systems. Tunneling is just one of the phenomena amenable
to semiclassical methods. Many problems, from rotating molecules to
many-body aspects of nuclear structure, can be modelled in terms of a
large spin, for which the semiclassical approximation is the natural
one \cite{lmgetc}. It is perhaps not widely realized that this
limit is not as well understood for spin as it is for massive
particles. The correct semiclassical spin propagator has been
recognized only gradually since the late 1980's \cite{skvs}, and its
consistency under composition of successive propagators was
shown only recently \cite{spg}. 

We begin by describing our specific problem. The molecular ion
[(tacn)$_6$Fe$_8$O$_2$(OH)$_{12}$]$^{8+}$ (or just \Fe8 for short)
forms a solid in which the \Fe8 groups are essentially noninteracting,
and each behaves like a single spin of magnitude $J=10$ in its ground
manifold. The degeneracy of the 21 Zeeman levels is partly lifted by
spin-orbit effects. Including an external magnetic field, the system is
well described by the anisotropy Hamiltonian \cite{cons}
\beq
 \ham = k_1 J_z^2  + k_2 J_y^2 - C [J_{+}^4 + J_{-}^4]
         - g \mu _B J_z H,
 \label{ham}
\eeq 
where $\bJ = (J_x, J_y, J_z)$ is a spin operator,
and $J_\pm = J_z \pm i J_y$ (not $J_x \pm i J_y$).
If $H = 0$, the spin has degenerate classical minima along the
$\pm\xhat$ axes, which cant symmetrically toward $\zhat$ as $H$ is
turned on. We wish to study the splitting
$\Dta$ due to quantum tunneling between these minima, particularly
as a function of $H$. It is convenient to work in reduced variables
$\lam = k_2/k_1$, $\lam_2 = CJ^2/k_1$, and $h = H/H_c$, with
$H_c = 2k_1 J/g\mu_B$. Measurements yield $g\simeq 2$,
$k_1 \simeq 0.338$~K, $k_2\simeq 0.246$~K, and $C \simeq 29\,\mu$K
\cite{barra,werns}.

The splitting $\Dta(H)$ oscillates 
with $H$, being quenched at certain $H$ \cite{gargepl,ersin}. The effect
can be viewed in terms of two instantons that
wind about the hard axis, $\zhat$, in opposite 
directions, and interfere with one another \cite{wilk}. Their actions
differ by $iJ{\cal A}$, where ${\cal A}$ is a real Berry phase equal to
the area of the closed loop formed by the two paths on the
complexified unit sphere. $\Dta(H) = 0$ whenever
$J {\cal A}(H)$ is an odd multiple of $\pi$. If $C=0$, there are
$2J$ quenching fields, at $H = (J -n -1/2)\Dta H$,
with $n = 0, 1, \ldots, 2J-1$,
and $\Dta H = (1-\lam)^{1/2} H_c/J = 0.263$~T.

\begin{figure}
\includegraphics{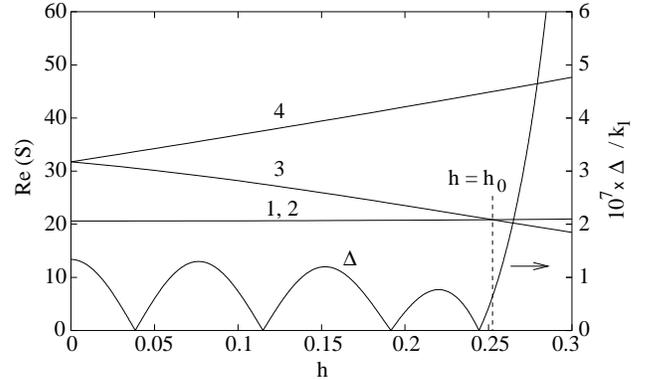}
\caption{\label{actionfigure} ${\rm Re\ }S$ for the \Fe8 instantons,
marked 1--4 as shown. $S_3$ and $S_4$ are purely real. Also shown
is the ground pair splitting $\Dta(h)$.}
\end{figure}

Turning on even a small $C$ has a big quantitative effect. With the
value for \Fe8, e.g., direct numerical diagonalization of $\ham$
(see \fno{actionfigure})
reveals only 4 quenching points for $H > 0$ instead of 10. 
The spacing remains quite regular, but is increased by $\sim50$\%.
Given the visual appeal of the interfering instanton picture, it is
interesting to ask how this picture is modified when $C\neq 0$.
To this end, we review some aspects of SU(2) instantons.

In the instanton method, one seeks a propagator such as
$K_{fi} = \mel{z_f}{\exp[-\ham T]}{z_i}$ via the path integral
\beq
K_{fi} = \int d[z] d[\baz] e^{-S(\baz(t),z(t))} \label{Kfipi}
\eeq
in the limit $T\to\infty$. Here, $\ket{z_{i,f}}$ are (unnormalized)
spin coherent states defined for any complex number $z$ by
$\ket{z} = e^{zJ_-} \ket{J,J}$, where $\ket{J,J}$ is the eigenstate of
$\bJ^2$ and $J_z$ with eigenvalues $J(J+1)$ and $J$. Further,
$\bJ\cdot\nhat\ket{z} = J\ket{z}$, if $z = \tan(\tta/2)e^{i\phi}$,
$(\tta,\phi)$ being the spherical polar coordinates of $\nhat$. We will
take the points $z_i$ and $z_f$ to be degenerate minima of the classical
energy. $S$ is the action for a path specified by $z(t)$
and $\baz(t)$, and is given by
\beq
S = -\int_{-T/2}^{T/2}
             \left[ J {\dot{\baz} z - \baz{\dot z} \over 1 + \baz z}
                           - H(\baz, z) \right] dt.
      \label{sk+d}
\eeq
Here, $z$ and $\baz$ are formal complex conjugates, but should be
regarded as independent variables, since both are
needed to determine $\tta$ and $\phi$, e.g. Further,
\beq
H({\bar z}', z) = \mel{z'}{\ham}{z}/ \tran{z'}{z}.
\eeq
For $\ham$'s such as (\ref{ham}) that are polynomial in $J_i$,
$H(\baz', z)$ is holomorphic in $z$ and antiholomorphic in $z'$ \cite{fn}.
In \eno{sk+d}, the
first term is the Wess-Zumino or Berry phase term. We will
refer to the two terms in $S$ as the kinetic and dynamical terms,
$S_K$ and $S_D$.

Instantons are paths that start at $z_i$ and end at $z_f$, and obey the
Euler-Lagrange (EL) equations,
\beq
\dot\baz = {(1 +\baz z)^2 \over 2J} {\ptl H \over \ptl z}, \qquad
\dot z = -{(1 +\baz z)^2 \over 2J} {\ptl H \over \ptl\baz}. \label{ELzz}
\eeq
It is easily verified that along these paths energy is conserved,
i.e., $dH(\baz,z)/dt = 0$. Because of this, and because $z_i$ and $z_f$
are energy minima, one cannot find a solution
lying on the real unit sphere. We must allow $\baz$ and $z$ to be completely
independent complex variables. In other words, $\baz(t)$ need not
equal $z^*(t)$, where the star denotes the {\it true\/} complex
conjugate. However, one can always find an instanton with endpoints on the
real sphere. This is because, if we denote the minimum energy by $\emin$,
the instanton obeys $H(\baz(t),z(t)) = \emin$, and one clearly
has $H(z^*_i,z_i) = H(z^*_f,z_f) = E_{\rm min}$.

Let us illustrate this using \eno{ham} with $C=0$. Then,
\beq
H(\baz,z) = k_1 J^2
            \lf[(1-\baz z)^2 - \lam (z - \baz)^2 - 2h (1 - \baz^2 z^2)
                   \over
                 (1 + \baz z)^2 \rt]. \label{hamfe8}
\eeq
The minima are at $\baz = z = \pm z_0$ where
$z_0 = [(1-h)/(1+h)]^{1/2}$, and $\emin = -k_1 J^2 h^2$. From
$H(\baz,z) = \emin$ we obtain $\baz(z)$:
\beq
\baz = {\rtl z \pm (1 - h) \over \rtl \pm (1 + h) z}. \label{zbar(z)}
\eeq
These equations give the instanton trajectories in $z$-$\baz$ space,
without giving the $t$ dependence. For general $z$, $\baz \ne z^*$, but
if $z=\pm z_0$, $\baz = z^*$. Thus the instanton endpoints are on the real
sphere. From \eno{zbar(z)}, we can now evaluate
$S$, and recover previous results, along with the interference
effect \cite{gargepl}.

If we now turn on $C$, the solutions (\ref{zbar(z)}) will evolve smoothly,
and continue to have classical end points. They will continue to interfere,
and one can find the fields where $\Dta$ vanishes by calculating
$J{\cal A}$ numerically. When we do this, we find that the
spacing agrees closely with the answer from direct diagonalization of
$\ham$, but we also find, incorrectly, quenching points at
$h > h_0 \simeq 0.25$.

The problem is that we have not formulated the principal of least action
(or Hamilton principal function, to be precise)
sufficiently carefully \cite{Fadd,Klau,spg}. One must in fact include an
explicit boundary term $S_B$ in $S$:
\beq
S_B = J\ln \lf[ {(1+\baz(-T/2) z_i) (1+\bzf z(T/2))
                 \over (1+ z^*_iz_i) (1+\bzf \bzf^*)} \rt].
   \label{Sbnd}
\eeq 
If we now vary $S = S_K + S_D + S_B$ {\it including the endpoints\/},
and set $\dta S$ to 0, we discover of course the EL equations
(\ref{ELzz}), but also that $\dta S$ has no terms in $\dta\baz(-T/2)$ and
$\dta z(T/2)$. This means that the boundary conditions on \eno{ELzz} are
\beq
z(-T/2) = z_i, \quad \baz(T/2) = \bzf, \label{bczz}
\eeq
and that $\bzi\equiv\baz(-T/2)$ and $z_f \equiv z(T/2)$ must be
left free. Otherwise, we would have four boundary
conditions on a second order system of differential equations, and the
problem would be overdetermined. The term $S_B$ can also be found
by careful time-slicing of the propagator. Its inclusion in $S$ has many
other nice consequences: e.g., the Hamilton-Jacobi equations
\beq
{\ptl \Scl \over \ptl \baz_f} = 2J {z_f \over 1 + \baz_f z_f}, \quad
{\ptl \Scl \over \ptl z_i} = 2J {\baz_i \over 1 + \baz_i z_i}.
         \label{HJ}
\eeq

Since $\bzi$ and $z_f$ are not determined, one may have solutions to
\etwo{ELzz}{bczz} with $\bzi \ne z^*_i$, $z_f \ne \bzf^*$.
These are the boundary jump instantons. Their
velocities $\dot z$ and $\dot\baz$ do not vanish at the end points
because the derivatives $\ptl H/\ptl z$ and $\ptl H/\ptl \baz$ are
not zero. Thus the instanton duration is finite, and although the
energy $E = H(\baz, z)$ is still a constant of motion, the value of
$E$ is not immediately obvious. In fact, since $S_D = \int H dt$,
we must choose $E = \emin$. Otherwise, when we sum multiinstanton
terms, the instantons with jumps will trivially dominate or be dominated
by those without jumps. This point comes out more easily in Klauder's
formulation. He argues that since the  continuum path integral is a
formal construct with meaning only as a limit of
its discrete version, one may add a term to the integrand for $S$ that is
quadratic in the velocities $\dot\baz$ and $\dot z$, with an infinitesimal
coefficient $\eps$ that is sent to $0$ at the end. The EL equations
are then a fourth order system, and one may specify all four $z_i$,
$\bzi$, $\baz_f$ and $z_f$. The instantons with jumps then appear
as solutions to the EL equations with internal boundary layers of thickness
$O(\eps)$ since the terms in $\ddot z$ and $\ddot\baz$ have coefficients
$\eps$.  Energy is conserved in these boundary layers too, and when one
takes the $\eps\to 0$ limit, they yield a contribution that is
explicitly independent of $\eps$ (which makes the procedure legitimate),
and is precisely equal to $S_B$ above. Note that $S_B = 0$ for an
instanton without jumps.

Hence the general procedure for finding all instantons is as
follows. For any $\ham$ with degenerate minima, we first find $\emin$,
and the classical minima $(z^*_i,z_i)$, $(z^*_f,z_f)$. We then find the
allowed values of $\bzi$ by solving
\beq
H(\baz,z_i) = \emin. \label{bzi}
\eeq
This equation has a double root at $\baz = z^*_i$, since
\beq
\left.{\ptl \over \ptl {\bar z}} H({\bar z}, z)
                                    \right|_{z^*_i,z_i}
 = \ \ 
\left. {\ptl \over \ptl z} H({\bar z}, z)
                                    \right|_{z^*_i,z_i}
 = \ \ 0. \label{Hanal}
\eeq
However, it may also have additional roots at $\baz\ne z^*_i$,
which will then be the end points of instantons with boundary jumps.
(A completely analogous procedure applies to $z_f$.) We then obtain
$\baz(z)$ for all instantons from energy conservation, making sure
that they connect on to the appropriate end points. This is enough
to compute $S_K$ and $S_B$ for each instanton (the time dependence is
not needed), and $S_D = \emin T$ for all of them. If we label the various
instantons by $\al$, we can write
\beq
\Dta = \sum_{\al} \gam_{\al} e^{-S_{\al}},
\eeq
where $\gam_{\al}$ is the prefactor arising from integrating over Gaussian
fluctations about each instanton. On physical grounds we expect
$\gam_{\al}$ to be of the same order for all $\al$ for smooth Hamiltonians, and
it may be estimated as the small oscillation frequency about the minimum.
(For instantons without boundary jumps, Ref. \cite{kspg} formulates how to
find $\gam_{\al}$.) Hence, the relative importance of various instantons is
determined largely by the actions $S_{\al}$.

\begin{figure}
\includegraphics{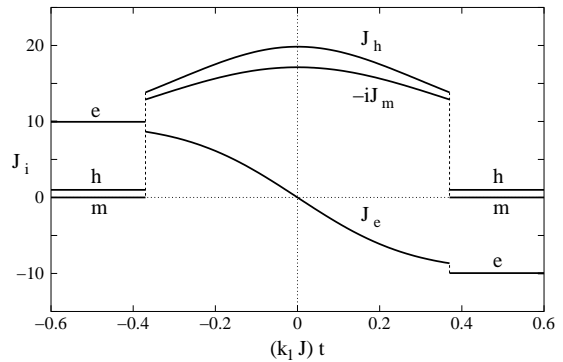}
\caption{\label{jumpinst}A jump instanton (number 3), for $H = 0.1H_c$.
Suffixes {\it e\/}, {\it m\/}, and {\it h\/} denote easy, medium, and
hard axes.}
\end{figure}

\begin{figure}[b]
\includegraphics{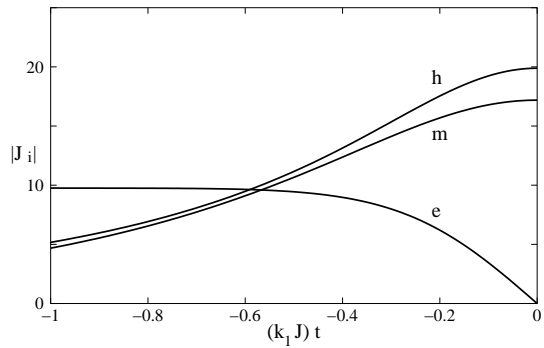}
\caption{\label{nojumpinst}Instantons without jumps for $H=0.1 H_c$.
We show $|J_i|$, since now all $J_i$ are complex. All $|J_i|$ are
even about $t=0$, and $J_m$ slowly asymptotes to 0 as $t \to \pm\infty$.}
\end{figure}

Let us now return to our model (\ref{ham}). It may be verified that
$H(\baz, z) = P(\baz, z)/(1+\baz z)^4$, where $P$ is a polynomial of
degree 4 in $\baz$ and also in $z$. Thus \eno{bzi} is a quartic in
$\baz$. Two of its roots are indeed $z_i^*$, and connect on to instantons
without jumps, but two are different and distinct and connect to
instantons with jumps. The equation $H(\baz,z) = \emin$ is also a quartic
and the solution $\baz(z)$ has four branches, correpsonding to the
different instantons. We label the first two instantons, which have
${\rm Im\ }S \ne 0$, and interfere with each other, 1 and 2, and the last
two, which have jumps and ${\rm Im\ }S = 0$, 3 and 4.
An instanton with a jump at one end also has a jump at the other end. 
A 180$^\circ$ rotation about $\zhat$ sends instanton 1 into 2, which
guarantees
${\rm Re}\,S_1 = {\rm Re}\,S_2$, $\gam_1 = \gam_2$. We show instanton 3
in Fig.~\ref{jumpinst}, 1 and 2 in Fig.~\ref{nojumpinst} (all for
$h = 0.1$), and ${\rm Re}\,S_{\al}(h)$ in Fig.~\ref{actionfigure}.
For any $h$, the dominant instanton is that with the least
${\rm Re}\,S$. Hence, except in the immediate vicinity of $h_0$, only
instantons 1 and 2 are relevant for $h < h_0$, and only 3 is relevant
for $h > h_0$. This explains why $\Dta$ does not oscillate for
$h > h_0$.

We can find the quenching fields numerically,
but we have also found an analytic approximation, based
on the small parameter $\zeta \equiv 4\lam_2 h^2$, which explains why
they are so regularly spaced. This result may also be of wider interest,
since $\Dta$ oscillations have now been seen in another system \cite{ww01}.
To derive this, it is better to use polar coordinates.
With $u \equiv \cos\tta$ and $s \equiv \sin\phi$,
the energy conservation condition takes the form
\beq
g(u,s) = -\tshf Z(s) u^4 + R(s) u^2 - 2h u + W(s) = 0, \label{gus}
\eeq
where
\bea
Z(s) &=& 4\lam_2 (1 + 6s^2 + s^4), \\
R(s) &=& 1 - \lam s^2 + 12 \lam_2 s^2 + 4\lam_2 s^4, \\
W(s) &=& g_0 + h^2 + \lam s^2 - 2\lam_2 s^4,
\eea
with $g_0 = -(\lam + h^2) -\emin \approx 2\lam_2 h^4$. $g(u,s)$
has four roots $u(s)$. Regarding $s$ as real, we are interested in the
complex conjugate pair of roots which tend to the energy minima
$\tta=\tta_0$, as $s \to 0$. Let the real and imaginary parts of these
roots be $A(s)$ and $B(s)$, i.e., let
\beq
u(s) = A(s) \pm i B(s).
\eeq
From the imaginary part of \eno{gus} we get
\beq
B^2 = A^2 - (R/Z) + (h/AZ),
\eeq
and if we substitute this result for $B$ into the real part, we get an
equation for $A$ alone:
\beq
4Z^2 A^6 - 4RZ A^4 + (R^2 + 2WZ) A^2 - h^2 = 0.
\eeq
We now make the self-consistently verifiable assumption that $A = O(h)$.
Then the terms $ZA^4$ and $Z^2A^6$ are $O(\zeta)$ and $O(\zeta^2)$
relative to the the remaining terms, and may be dropped. This yields
$A = h(R^2 + 2WZ)^{-1/2}$. The quantity $R^2 + 2WZ$ can be seen to be
a fourth order polynomial in $s$, and depends on $h$ only through the
combination $\lam_2(h^2 + g_0)$, which is $O(\zeta)$. If we neglect this
weak $h$ dependence, we get
\bea
A(\phi) &\approx& h(1 + P_2 \sin^2\phi + P_4 \sin^4\phi)^{-1/2},
                      \label{sola2}\\
P_2 &=& -2\lam + 24 \lam_2 + 8 \lam_2 \lam,  \\
P_4 &=& \lam^2 + 8 \lam_2 + 24\lam\lam_2 + 128 \lam_2^2.
\eea 
Since $S_K = iJ\int (1-\cos\tta)\dot\phi\, dt$ in $\tta$, $\phi$ variables,
${\cal A} = 2\pi - \int_0^{2\pi} A(\phi)d\phi$. If we keep only instantons
1 and 2, $\Dta = 2\gam_1 \exp(-{\rm Re}S_1) \cos(J{\cal A}/2)$ up to
a phase factor. Using \eno{sola2}, we see that the zeros of $\Dta$ are
equally spaced with spacing $\Dta H = \pi H_c / J I(\lam,\lam_2)$, where
%
\beq
I(\lam,\lam_2) =
      \int_0^{\pi}\frac {d\phi}{ (1 + P_2 \sin^2\phi + P_4 \sin^4\phi)^{1/2}}.
\label{perint}
\eeq
For the \Fe8 parameters, $I=3.88$, implying $\Dta H = 0.409$~T.
The experimental value is 0.41~T.

We conclude with some general remarks about instantons with boundary
jumps (or internal boundary layers, in the Klauderian view).
It is clear that they must be present in all coherent state path
integrals, not just for spin, and our discussion is easily extended to
these cases. It
would be interesting to find other concrete instances where they occur,
both in quantum mechanics, and in field theories. It would also be
interesting to reaxamine problems such as a particle in a one dimensional
potential well in a coherent state formulation.

%
%

\begin{thebibliography}{99}
\bibitem{mis1} It should be stressed that our remarks apply only to
{\it tunneling\/}, i.e., to infinite time propagators. Paths with
nonclassical endpoints are known to be important for finite time problems,
e.g., Larmor precession in the semiclassical approximation.
In these cases, however, these are the {\it only\/}
paths that contribute. There is no problem of which we are aware in
which both types of paths, with and without jumps, need to be included.
\bibitem{cons} In this paper we only consider external fields along
$\zhat$. Also, note that one can add a term proportional to $\bJ\cdot\bJ$
to $\ham$, since that commutes with
$\ham$. This can change the superficial appearance of $\ham$ and the
coefficients of $J_x^2$, etc. The physics, of course, is unaffected.
\bibitem{fn} See, e.g., \eno{hamfe8}. This is a key benefit of using
$z$ and $\baz$. By enforcing analyticity
in various approximations, many calculations can be streamlined.
Explicit model calculations, however, are often simpler in $\tta$ and
$\phi$.
\end{thebibliography}
\end{document}